\documentclass[12pt]{article}
\usepackage{multirow}
\usepackage{amsmath,amsfonts,amssymb,bm,slashed,bbm,mathrsfs}
\usepackage{graphicx,color}
\usepackage{graphicx}
\graphicspath{{./Pic/}}

\newcommand{\ep}{\epsilon}

\begin{document}
\title{Induced color charges in QGP at Polyakov's loop and chromomagnetic fields}
\author{
  V. Skalozub\thanks{e-mail: Skalozub@ffeks.dnu.edu.ua}, ~~I. Gamolsky  \\
{\small Oles Honchar Dnipro National University, 49010 Dnipro, Ukraine}}
\date{ June 24, 2023}
\date{\small}
\maketitle
\begin{abstract}
In quark-gluon plasma (QGP), at high temperature the spontaneous generation of color magnetic fields, $H^3(T), H^8(T)  \not = 0$, and usual magnetic field $H(T)  \not = 0$ is realized. Also,   classical field - $A_0(T)$ condensate  directly related to the Polyakov loop - is  spontaneously created.
 The common generation  of both of them within  the two loop effective potential was investigated recently for SU(2) gluodynamics in  \cite{skal21-18-738}, \cite{bord22-82-390}. The values of the field strengths and the mechanism of the magnetic fields stabilization due to $A_0(T)$  have been discovered.

In the present paper, we develop further these investigations in order to find new effects  of QGP. First of all, we generalize the $SU(2)$ consideration to the full QCD case.
Then we calculate the induced color charge $Q_{ind}^3$   at the background magnetic field
$ H(T)  \not = 0$ and the condensates $A_0^3(T)$. The  $ A_0^8(T) $ is zero in used two-loop approximation. So, $Q_{ind}^8$ is zero  also.
 We conclude that magnetic field essentially influences the induced charge compared to the zero field case investigated in  \cite{skal21-cond-156}.
 The extension to other type magnetic fields is given.
	
Key words: spontaneous magnetization, high temperature, asymptotic freedom, effective potential, $A_0$ condensate, effective charge, effective vertexes.
\end{abstract}
\section{Introduction}
Quark-gluon plasma (QGP) is the  state of  matter consisting   of quarks and gluons liberated form nuclei at high temperature due to  asymptotic freedom of non-Abelian gauge fields.  A classical background of QGP is formed out of two type  condensates - the $A_0(T)$ (Polyakov loops (PL)) and  the spontaneously generated temperature dependent chromomagnetic $H^3(T), H^8(T)$, where 3 and 8  are color indexes of $SU(3)$ group,   and usual magnetic $H(T) $ fields. The first kind condensate   results  in the color $Z(3)$ symmetry breaking and the Furry's theorem violation. The second ones considerably change the spectra of quarks and gluons as well. So, new phenomena have to be realized. In particular, the induced color charges $Q^3_{ind}, Q^8_{ind}$ are expected. The PL and $A_0(T)$ are the order parameters for the phase transition.    At low temperature, PL and $ A_0$ equal zero. At high temperature they become nonzero. The same concerns the magnetic fields.  Recently, on the principles of the Nielsen's identity method and new type integral and sum representations  we derived the gauge invariant expression for the $ A_0$ condensate in the magnetic fields in two-loop approximation  \cite{skal21-18-738}, \cite{bord22-82-390}. This (in particular)  opens a possibility for calculating the induced color charges for this case.

 To realize that, we have  to calculate the contribution of the diagram depicted in Figure 1. Here, the solid line presents the quark propagator in the $A_0$ and magnetic fields and the wavy line presents the zero component of gluon fields $ G_0^3$ or $G_0^8$. At finite temperature, in the Matsubara formalism one has to calculate the temperature sum  over discrete momenta $p_4 = 2 \pi T (l + 1/2), l = 0, \pm 1,  ...$, integrate over momentum component $p_3$ oriented along the space field direction, calculate the sum over spin variable  $\sigma = \pm 1$  and  sum up over n = 0, 1, 2,... , in correspondence to the fermion spectrum in magnetic field H: $(p_4 + g A_0)^2 = m^2 + p_3^2 + (2 n + 1) g H - \sigma g H $. Here we write  $g H$ as a general expression corresponding to each of the fields. For instance, this is $e H$ for usual magnetic field, $g H^3, g H^8$ - for color fields.

  In what follows, we apply  the low level approximation, $n = 0, \sigma = + 1$ giving a leading contribution for strong external fields.
We  obtain that the induced color charge $Q^3_{ind}$ is nonzero. The presence of the magnetic fields changes the values of it compared to the zero field case. So that the QGP has to be magnetized and  color charged.

\begin{figure}[h]
\begin{center}
\includegraphics[width=0.25\textwidth]{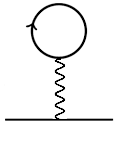}
\caption{Tadpole diagram}
\label{tadpole}
\end{center}
\end{figure}
\section{Induced Color charge}
In this section, we calculate the induced color charge generated by the tadpole diagram of Fig. 1. In charged basis, we have two components of the induced charge for the shifts $A_0^3$ and $A_0^8$. But accounting  for the result \cite{skal21-cond-156} $A_0^8 = 0$ ,  we have to calculate the contribution for the case $(A_0)_\mu^a=A_0\delta_{\mu4}\delta^{a3}$. The explicit form in the Euclid space-time   is $Q_4^3 Q^3_{ind} $, and we have

\begin{equation}
\label{Q3_init}
    Q_{ind}^3=  \frac{g}{\beta}\sum_{p_4}\int\frac{d^3p}{(2\pi)^3}Tr\left[\frac{\lambda^3}{2}\gamma_4\frac{\hat p_\sigma\gamma_\sigma+m}{\hat p^2 +m^2 }\right],
\end{equation}

\noindent where $\hat p=(p_4=p_4\pm A_0,\mathbf{p}),\,p_4=2\pi T(l+1/2),\,l=0,\pm1,\dots$. The trace is calculated over either space-time or color variables. $\lambda^3$ is Gell-Mann matrix.  Here also  we noted as $ A_0 $ the value $ A_0 = \frac{g A_0}{2}$.

Calculating the  traces over the space and the internal indexes we get,

\begin{equation}
\label{DWI_Q3}
    Q_{ind}^3 = \frac{4 g}{\beta} \int\frac{d^3p}{(2\pi)^3}\sum_{p_4}\frac{(p_4+A_0)}{(p_4+A_0)^2 + \ep_p^2},
\end{equation}
where $\ep_p^2 = \vec{p}^2 + m^2$. In case of nonzero field,
$\ep_p^2 = {p_3^2 + m^2 + (2n + 1) gH - g H \sigma} $.

To calculate the temperature sum we use the following representation

\begin{equation}
    Q_{ind}^3= 4 g \int\frac{d^3 p}{(2\pi)^3}\frac{\beta}{\pi} \oint_{C}\tan\frac{\beta\omega}{2}\frac{(\omega+A_0)}{(\omega+A_0)^2+\ep^2_p}d\omega.
\end{equation}
This is in the case of zero field. If the field is nonzero, we have to replace $\frac{d^3 p}{(2 \pi)^3} -> \frac{d p_3}{2 \pi}  \frac{gH }{(2 \pi)^2}$ in correspondence to the particle spectrum.  The integrand function has two imaginary poles of the first order. We use the residues to find the charge value.

The result, after transformation into spherical coordinates and angular  integration, is

\begin{equation}
    Q_{ind}^3= \frac{g \sin{(A_0 \beta)}}{\pi^2} \int_0^{\infty} p^2 d p \frac{1}{\cos\beta A_0 +\cosh ( \beta \ep_p ) }.
\end{equation}

In what follows, we calculate the integral in the high-temperature limit $T\to\infty$. In this case we use


\begin{equation}
  \ep_p  =\sqrt{\mathbf{p}^2+m^2}\approx|\mathbf{p}| + \frac{1}{2}\frac{ m^2}{|\mathbf{p}|}
\end{equation}
because large values of momentum give dominant contribution.

 After integration over momentum we obtain  at zero field \cite{skal21-cond-156}
\begin{equation}\label{Q3_fin}
    Q_{ind}^3= g A_0^3 \bigl[ \frac{T^2}{3} - \frac{m^2}{2 \pi^2}\bigr].
\end{equation}
As we see,  the first term  depends on temperature as $\sim T^2$. The second one depends on mass, only.  At high-temperature,  the first term is dominant and  the plasma acquires the spontaneous induced charge in the case $m=0$, also.

Now, we turn to nonzero $H$.  Using the low Landau level approximation, $\sigma = + 1, n = 0$, we get after integration over $p_3$ momentum
\begin{equation}  Q_{ind}^3(H, T) = g \frac{g H}{2 \pi^3} \frac{\sin (A_0^3 \beta)}{\beta } ( 1 + 7 \beta^2 m^2 Zeta^{'} (- 2) ).
\end{equation}
 At high temperature, this expression reads
\begin{equation} \label{QHT} Q_{ind}^3(H, T) =  \frac{g H}{2 \pi^3}  (g A_0^3 ) ( 1 + 7 \beta^2 m^2 Zeta^{'} (- 2) ).
\end{equation}
Note that numerically $Zeta^{'} (- 2) = - 0.03044485$.
Thus, one of the consequences of the $A_0$ condensate presence is the $Z(3)$ symmetry and the $C$-parity violation, which leads to the induction of color charge in the plasma.

As we noted in Introduction, the factor $g H$ presents different kinds of magnetic fields - usual magnetic field $eH$, color magnetic fields $gH^3, gH^8$ or even some combination of  them. Remind also that in the above formulas $A_0^3$ has to be substituted by $\frac{g A_0^3}{2}.$
\section{Discussion}
Let us compare the values of induced color charges given by formulas (\ref{Q3_fin}) and (\ref{QHT}). The first leading in temperature terms are of interest, now. Both expressions have the factor $g A_0^3$ but a different temperature dependence. In former case, the factor $\sim T^2$ stands and determines high temperature behavior. In latter case, it is determined by the temperature dependence of the magnetic field $H(T)$. This behavior has been investigated recently in two-loop approximation \cite{skal23-2305.00757}
\begin{equation}
\label{bmin} g H  = \frac{1}{16}\frac{(2 a_2 \alpha_s)^{2/3}}{\pi^{2/3}} T^2.
\end{equation}
Here, $a_2 = 1.856$ is number, $\alpha_s = g^2/(1 + \frac{11}{12} \frac{g^2}{\pi^2} \log (  T/\mu))$ is running coupling constant, $\mu $ is a normalization point for temperature.  So, at reference temperature $\alpha_s = g^2$.  Because of this factor, the value of the field strength is always smaller compared to  $T^2$. As a result, the induced color charge (\ref{QHT}) in the magnetic field is also smaller compered to eq.(\ref{Q3_fin}).

On the other hand, during carried out calculations we have taken the field strength as a given number which is arbitrary. So that it can be the field produced  by some external current. In this case the induced color charge will be completely determined by external field. Such a situation is expected and discussed for heavy ion collision experiments.


\end{document}